\documentclass[11pt,twoside]{article}

\usepackage{asp2004}
\usepackage{epsf}

\newcommand{\lam}{$\lambda$}

\begin{document}
\title{Studying Transition Region Phenomena with Solar-B/EIS}   
\author{}   
\affil{}    

\begin{abstract} 
Transition region lines in active regions can become strongly enhanced
in coronal footpoints and active region blinkers. The weak transition
region lines found in the Solar-B/EIS wavebands will thus become
useful for diagnostic studies of these events. EIS count rates predicted
from SOHO/CDS spectra are presented, and a \ion{Mg}{vii} density
diagnostic is highlighted.
\end{abstract}



\section{Introduction}

The EUV Imaging Spectrograph (EIS) on Solar-B will provide high
resolution EUV spectra over the wavelength ranges 170--210~\AA\ and
250--290~\AA\ where there are many strong emission lines from coronal
species, particularly the iron ions \ion{Fe}{x--xvi}. Strong
transition region lines in UV are found only above 400~\AA, however,
and thus only weak lines are found in the EIS wavebands. 
Table~\ref{tbl.transitions} lists the most important of these lines
together with their temperature of maximum abundance ($T_{\rm max}$).
The following sections use data obtained with the Coronal Diagnostic
Spectrometer (CDS) on SOHO to show that these transition region lines
can become strong in certain active region features and will
complement the coronal line data.

\begin{table}[h]
\caption{Key transition region lines that will be observed by EIS.}
\smallskip
\begin{center}
{\small
\begin{tabular}{llll}
\tableline
\noalign{\smallskip}
Ion &Wavelength/\AA &Transition &Log\,$T_{\rm max}$/K\\
\tableline
\noalign{\smallskip}
\ion{Mg}{v} &276.58 
    &$2s^22p^4$ $^1D_2$ -- $2s2p^5$ $^1P_1$ &5.5 \\
\ion{Fe}{viii} &185.12 
    &$3p^63d$ $^2D_{5/2}$ -- $3p^53d^2$ $^1P_{1}$ &5.6\\
\ion{Mg}{vi} &268.99
    &$2s^22p^3$ $^2D_{3/2}$ -- $2s2p^4$ $^2P_{1/2}$ &5.7\\
&270.40
    &$2s^22p^3$ $^2D_{3/2,5/2}$ -- $2s2p^4$ $^2P_{3/2}$ &5.7\\
\ion{Mg}{vii} &278.39
    &$2s^22p^2$ $^3P_2$ -- $2s2p^3$ $^3S_1$ & 5.8 \\
&280.75
    &$2s^22p^2$ $^1D_2$ -- $2s2p^3$ $^1P_1$ & 5.8 \\
\ion{Si}{vii} &275.35
    &$2s^22p^4$ $^3P_2$ -- $2s2p^5$ $^3P_2$ &5.8 \\
\tableline
\end{tabular}
}
\end{center}
\label{tbl.transitions}
\end{table}

\section{Density diagnostics}

\begin{figure}[t]
\plottwo{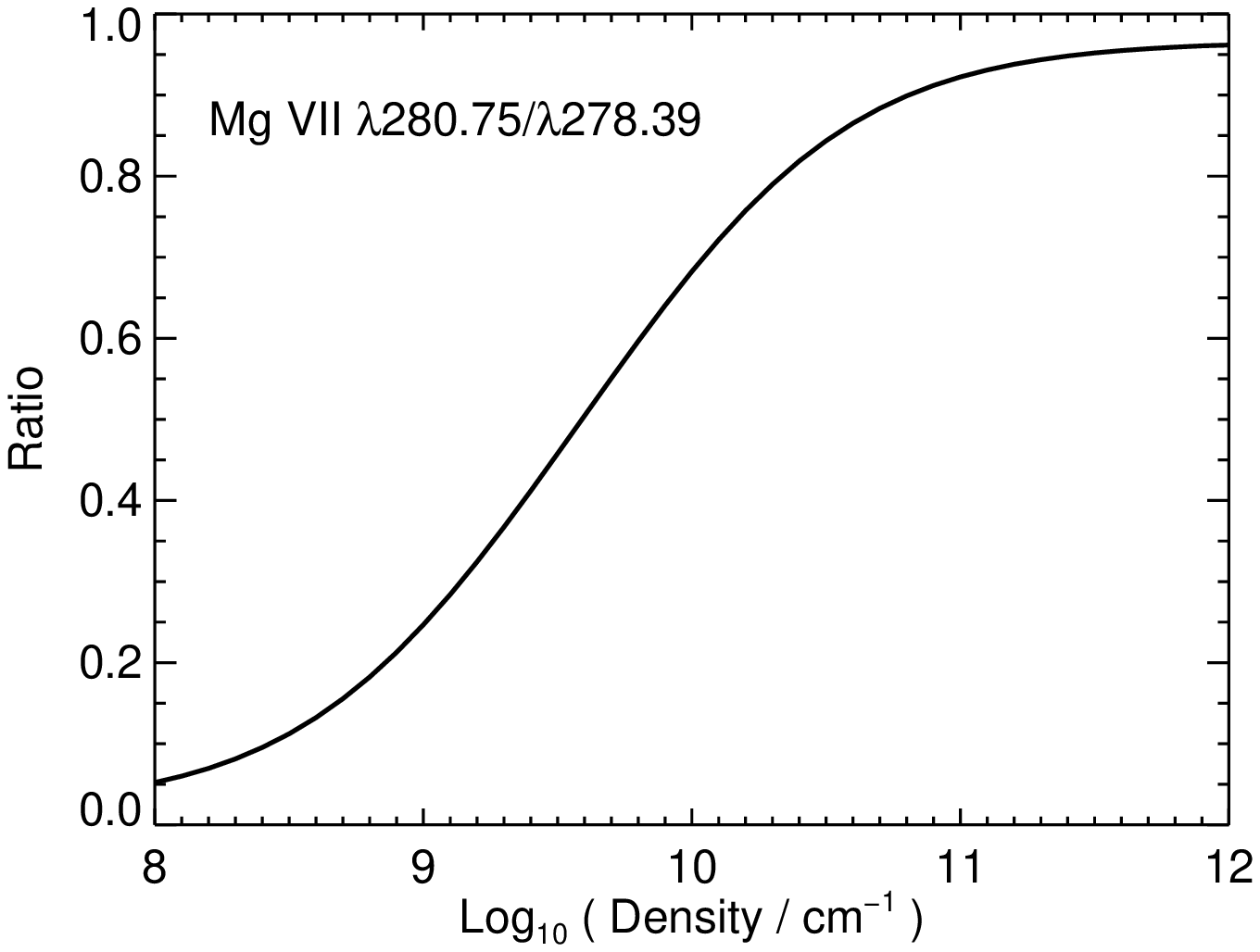}{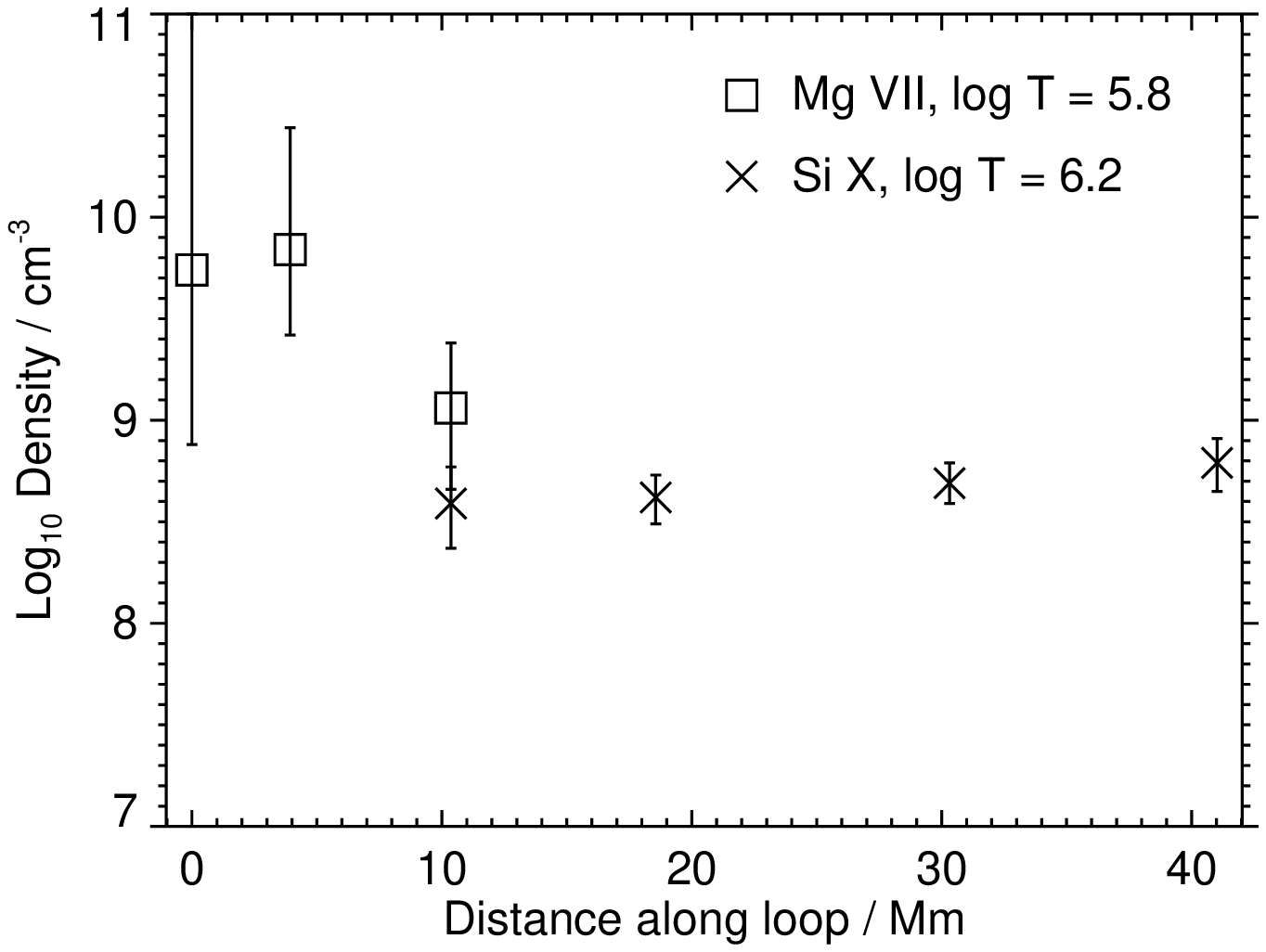}
\caption{The left-hand panel shows the density sensitivity of the
  \ion{Mg}{vii} 280.75/278.39 ratio. The right-hand panel shows
  density measurements from \ion{Mg}{vii} obtained with CDS coupled
  with \ion{Si}{x} density measurements, illustrating how
  \ion{Mg}{vii} is used to probe the footpoint regions of the coronal
  loop discussed in Sect.~\ref{sect.footpoints}.}
\label{fig.mg7-dens}
\end{figure}

EIS has excellent density diagnostic capability at coronal
temperatures, with several line ratio diagnostics with strong
sensitivity to density. There is also one diagnostic below 10$^6$~K:
\ion{Mg}{vii} 280.75/278.39 formed at $\log\,T =
5.8$. Figure~\ref{fig.mg7-dens} shows the 
sensitivity of the ratio as a function of density derived from v5.1 of
the CHIANTI database \citep{landi05}. Care has to be taken with
the \lam278.39 line as it is blended with \ion{Si}{vii} \lam278.44,
however this 
can be accounted for if the \ion{Si}{vii} \lam275.35 line is observed,
since the 
\ion{Si}{vii} \lam278.44/\lam275.35 ratio has a value of 0.32 in all solar
conditions. 

The value of the \ion{Mg}{vii} ratio is illustrated in
Figure~\ref{fig.mg7-dens}, where 
density measurements obtained with CDS in the coronal loop discussed
in Sect. 3 are presented. \ion{Mg}{vii} is used in the loop footpoint where
the coronal diagnostics are not useful.

\section{Coronal loop footpoints}\label{sect.footpoints}

\begin{table}[b]
\caption{CDS line intensities from the loop footpoint shown in
  Figure~\ref{fig.cds_images} are given for four magnesium ion
  lines. Using the CHIANTI database, the intensities for lines of the
  same ions that appear in the EIS wavebands are predicted, and the
  event rates at the detector are given. These are compared with average active region event
  rates.}
\smallskip
\begin{center}
{\small 
\begin{tabular}{llllllll}
\tableline
\noalign{\smallskip}
&\multicolumn{2}{c}{CDS lines}&
&\multicolumn{3}{c}{EIS lines}&Average \\
\cline{2-3}\cline{5-7}
\noalign{\smallskip}
Ion &\lam &Intensity &&\lam &Intensity &Signal &Signal \\
&[\AA] &[erg/cm$^2$/s/sr]&&[\AA]
  &[erg/cm$^2$/s/sr] &[DN/s] &[DN/s] \\
\tableline
\noalign{\smallskip}
\ion{Mg}{v}  &353.2 &561 &~&276.58 &309 &14.7 &1.1 \\
\ion{Mg}{vi} &349.2 &812 &&270.40 &487 &29.8 &6.6 \\
\ion{Mg}{vii}&367.7 &585 &&278.39 &450 &18.3 &3.3 \\
             &319.0 &602 &&280.75 &331 &10.5 &5.9 \\
\tableline 
\end{tabular}
}
\end{center}
\label{tbl.intensities}
\end{table}

Coronal loops observed with CDS often show bright emission in upper
transition region lines at their footpoints. They are particularly
strong in lines of \ion{Mg}{v}, \ion{Mg}{vi} and
\ion{Mg}{vii}. Emission lines from these 
ions are also seen with EIS, and we can directly estimate emission
line strengths for EIS based on the CDS spectra. 

\begin{figure}[t]
\centerline{\epsfxsize=1.8in\epsfbox{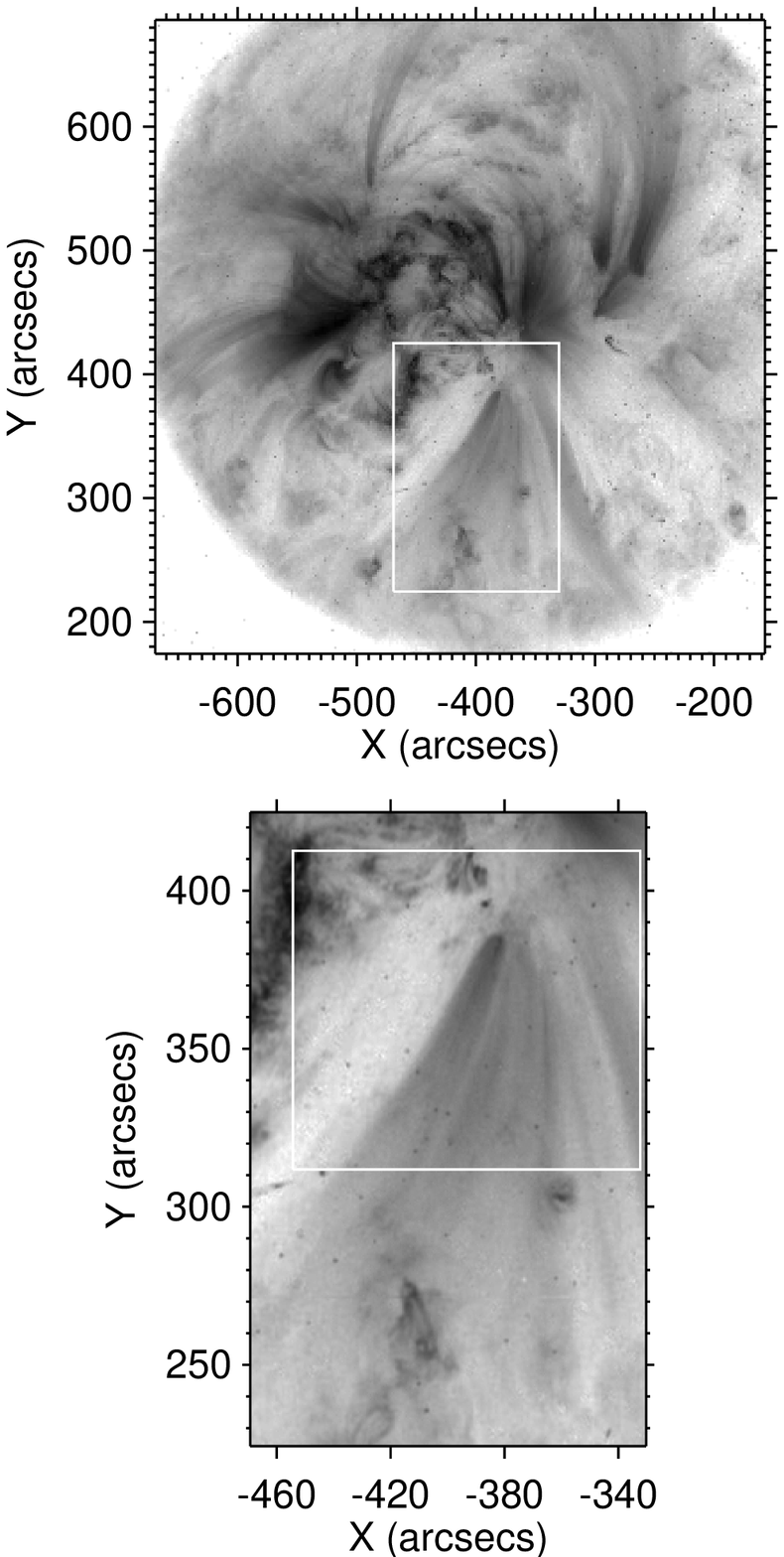}
            \epsfxsize=3.2in\epsfbox{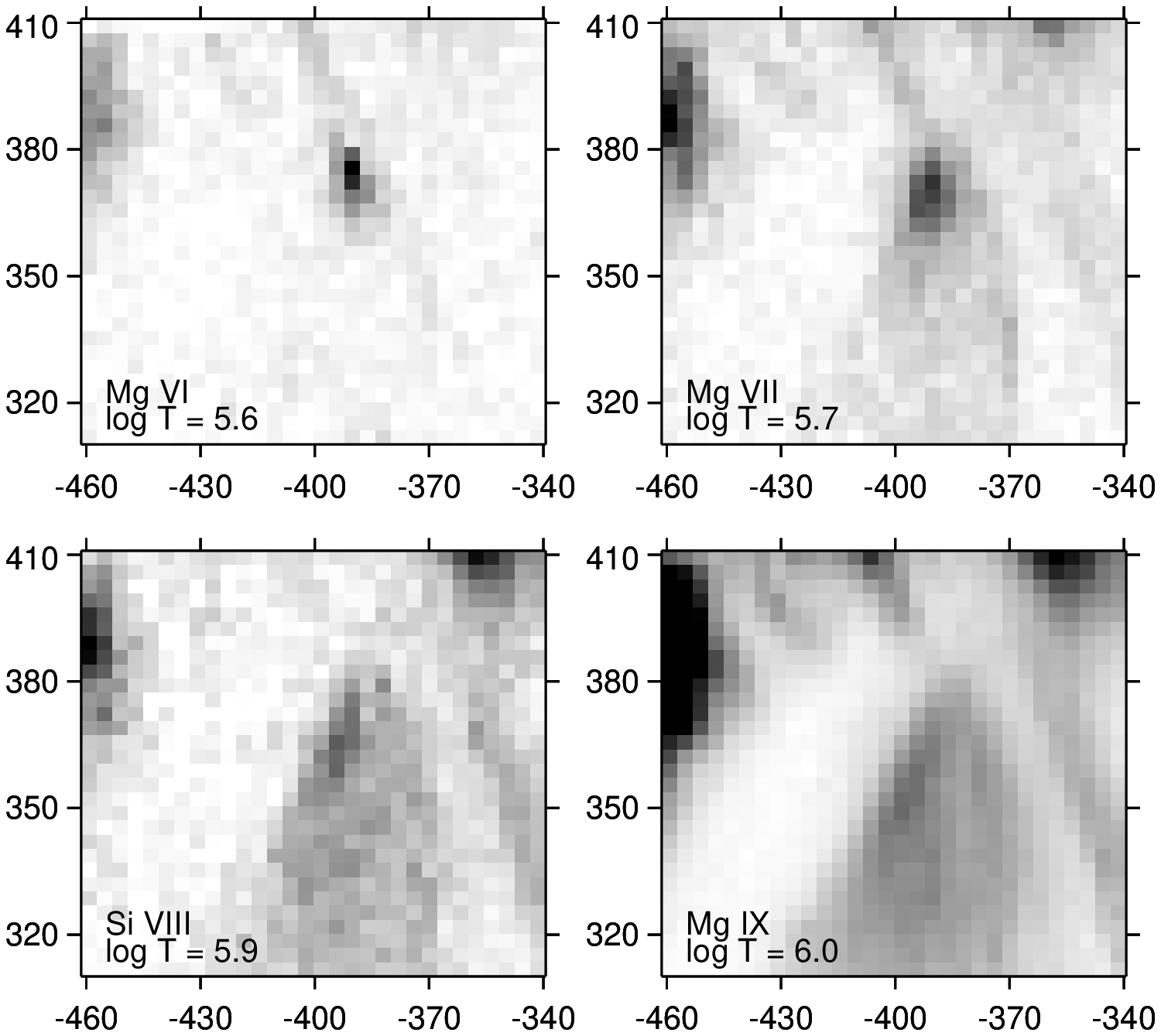}}
\caption{The upper left panel shows a TRACE 173 image of active region AR
            8227 observed on 1998 May 29, with the white box
            indicating the region shown in greater detail in the lower
            left panel. The white box in this image shows the region
            observed with CDS, from which four images are shown in the
            right hand panels. These four images show the loop at
            different temperatures, with the footpoint clearly seen in
            \ion{Mg}{vi} and \ion{Mg}{vii}. The images are shown as negatives with the most intense
  regions coloured black.} 
\label{fig.cds_images}
\end{figure}

The images in Figure~\ref{fig.cds_images} show one particular isolated
loop, where the 
CDS data show the footpoint to be bright in the cool Mg
ions. Intensities from CDS are given in Table~\ref{tbl.intensities},
together with 
intensities for the EIS lines predicted from the CDS lines using
atomic data from CHIANTI. The intensities are then converted to data
numbers (DN) per second, and compared with average active region
values showing the large enhancements found in the footpoints. 

\section{Active region blinkers}

\begin{figure}[h]
\plotone{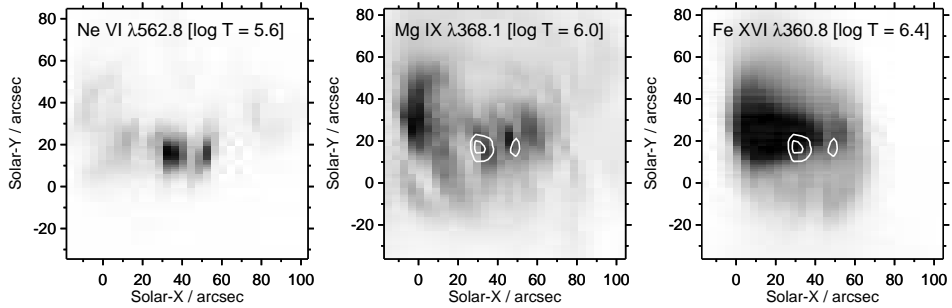}
\caption{Images of active region AR 7968 obtained on 1996 June 6 by
  CDS. The \ion{Ne}{vi} image shows two short-lived active region
  blinkers near the centre of the raster. The locations of these
  brightenings are superimposed on the two coronal line images through
  contours and show that they found in the central part of active
  regions. The images are shown as negatives with the most intense
  regions coloured black.}
\label{fig.blinkers}
\end{figure}

Blinkers are flashes in the transition region, first studied with CDS
(Harrison 1997). In active regions, the intensities of blinkers can be
factors 10$^3$--10$^5$ higher than in the quiet Sun and have densities up to
10$^{12}$ cm$^{-3}$ \citep{parnell02,young04}. They appear to be related
to new flux emergence and have photospheric abundances
\citep{young97}, being particularly pronounced in emission lines of Ne
and O. An 
example from CDS is shown in Figure~\ref{fig.blinkers}. 

As with loop footpoints, the \ion{Mg}{v--vii} lines are strongly enhanced in
these blinker events, however the blinkers are usually found in the
central part of active regions where coronal lines are strong. Studies
with CDS were thus compromised by significant blending with coronal
lines. This will not be a problem with the 
higher spectral resolution of EIS,  while
the higher spatial resolution of EIS will concentrate the blinkers
into 1 or 2 spatial pixels, leading to higher photons/pixel than found
with CDS. 

\section{Conclusions}

Weak transition region lines in the EIS wavelength bands will provide
extremely valuable science in studies of active regions. The coolest
ions (\ion{Mg}{v}, \ion{Mg}{vi}, \ion{Fe}{viii}) will highlight
coronal loop footpoints: of great benefit when combining with the
photospheric/chromospheric data from SOT. The \ion{Mg}{vii} density
diagnostic will provide a density measurement in these footpoint
regions. 

Active region blinkers will be seen in the transition region lines,
and the high spectral resolution of EIS will allow the relationship
with explosive events to be understood. Again by coupling with the SOT
data, the precise relationship between these blinkers and the magnetic
field evolution can be determined. 

Scientists are strongly encouraged to include one or more of the cool
lines discussed here when designing EIS studies.






\begin{thebibliography}{}

\bibitem[Landi et al.(2005)]{landi05}
  Landi, E., Del Zanna, G., Young, P.R., et al., 2005, 
  ApJS, in press 

\bibitem[Parnell et al.(2002)]{parnell02}
  Parnell, C. E., Bewsher, D., \& Harrison, R. A., 2002, 
  Sol. Phys., 206, 249

\bibitem[Young(2004)]{young04}
  Young, P. R. 2004, 
  Proc. of 13th SOHO Workshop, ESA SP-547, p.257

\bibitem[Young \& Mason(1997)]{young97}
  Young, P. R., \& Mason, H. E. 1997,
  Sol. Phys., 175, 523 

\end{thebibliography}
\end{document}